\documentclass{emulateapj}
\usepackage{graphicx}

\slugcomment{To appear in ApJ}
\shorttitle{A systematic search for BHBs}
\shortauthors{Tsalmantza et al.}

\def\Mgii{Mg\,{\sc ii}}

\def\Oiii{[O\,{\sc iii}]}
\def\Oii{[O\,{\sc ii}]}
\def\Nii{[N\,{\sc ii}]}
\def\Sii{[S\,{\sc ii}]}
\def\Ha{H$\alpha$}
\def\Hb{H$\beta$}

\def\kms{km s$^{-1}$}
\def\lsim{\mathrel{\rlap{\lower 3pt \hbox{$\sim$}} \raise 2.0pt \hbox{$<$}}}
\def\gsim{\mathrel{\rlap{\lower 3pt \hbox{$\sim$}} \raise 2.0pt \hbox{$>$}}}

\begin{document}

\title{
A systematic search for massive black hole binaries in SDSS spectroscopic sample
}

\author{P. Tsalmantza\altaffilmark{1}, R. Decarli\altaffilmark{1}, M. Dotti\altaffilmark{2,3}, David W. Hogg\altaffilmark{1,4} 
}
\altaffiltext{1}{Max-Planck Institut f\"{u}r Astronomie, K\"{o}nigstuhl 17, D-69117, Heidelberg, Germany. E-mail: {\sf vivitsal@mpia.de}, {\sf decarli@mpia.de}}
\altaffiltext{2}{Max-Planck-Institut f\"{u}r Astrophysik, Karl-Schwarzschild-Str. 1, D-85748, Garching, Germany.}
\altaffiltext{3}{Dipartimento di Fisica G.~Occhialini, Universit\`a degli Studi
  di Milano Bicocca, Piazza della Scienza 3, 20126 Milano, Italy}
\altaffiltext{4}{Center~for~Cosmology~and~Particle~Physics, Department~of~Physics, New~York~University, 4~Washington~Place, New~York, NY 10003, USA}

\begin{abstract}
We present the results of a systematic search for massive black hole
binaries in the Sloan Digital Sky Survey spectroscopic database.  We
focus on bound binaries, under the assumption that one of the black
holes is active. In this framework, the
broad lines associated to the accreting black hole are expected to
show systematic velocity shifts with respect to the narrow lines,
which trace the rest-frame of the galaxy.
For a sample of 54\,586 quasars and 3\,929 galaxies at redshifts
$0.1<z<1.5$ we brute-force model each spectrum as a mixture of two
quasars at two different redshifts.  The spectral model is a
data-driven dimensionality reduction of the SDSS quasar spectra based
on a matrix factorization. We identified 32 objects
with peculiar spectra. Nine of them can be interpreted as black
hole binaries. This doubles the number of known black hole binary 
candidates. We also report on the discovery of a new class
of extreme double-peaked emitters with exceptionally broad and faint
Balmer lines. For all the interesting sources, we present detailed
analysis of the spectra, and discuss possible interpretations. 
\end{abstract}
\keywords{methods: data analysis; methods: statisticals; 
quasars: general; quasars: emission lines}

\section{Introduction}

Massive black hole (BH) pairs are the natural outcome of mergers
through the hierarchical formation of galaxies. Examples of unbound BH
pairs, with separations of $\lsim 1$ kpc, have been observed, as in
the prototypical case of NGC 6240 \citep[][see Colpi \& Dotti 2009 for
  a recent review]{komossa2003}. At separations of $\sim$ few pc the
two BHs start experiencing their own gravitational interaction, binding in
a BH binary (BHB). Observing BHBs is challenging, since they
cannot be spatially resolved in optical and X-ray. The only spatially
resolved BHB candidate to date is hosted by the elliptical galaxy
0402+379 \citep{maness2004,rodriguez2006}. The two flat-spectrum
radio sources,
corresponding to the two components of the candidate BHB, have a
projected separation of $\approx 7$ pc. At the distance of 0402+379
($z = 0.055$) this corresponds to few milliarcsec, an angular scale
that can be probed only through radio interferometry. A second BHB
candidate is the BL Lac OJ287 \citep[see][and references
  therein]{valtonen2008}. It shows a $\approx 12$ yr modulation in its
light curve, that has been interpreted as related to the orbital
period of a BHB lurking in the nucleus of the AGN.

All the other BHB candidates discussed to date have been identified
by studying their optical and near--infrared spectra. In a BHB scenario,
the broad lines (BLs) emitted by gas bound to each BH may be
red- or blue-shifted with respect to their host galaxy redshift, as a
consequence of the Keplerian motion of the binary
\citep{begelman1980}. Furthermore, the BL region of each BH can be
perturbed and stripped by the gravitational potential of the
companion, resulting in peculiar flux ratios between BLs with
different ionization potential \citep{montuori2010}.  This
spectroscopic approach does not suffer any angular resolution
limitations: Actually, the closer (and more massive) the binary is,
the more shifted/deformed the BLs are.  Thanks to the existence of
large spectroscopic surveys, such as the Sloan Digital Sky Survey
(SDSS), a large region of the sky can be probed. 
Up to date five spectroscopically identified candidates have been
presented: J0927+2943
\citep{komossa08,bogdanovic09,dotti09}, J1536+0441
\citep{boroson09}, J1050+3456 \citep{shields09}, 4C+22.25
\citep[J1000+2233 in this paper,][]{decarli_4c2225}, and
J0932+0318 \citep{barrows11}. Such a small number of objects
is marginally compatible with the theoretically predicted number of
sub-parsec BHBs at $z \lsim 0.7$ \citep[5--10, given the merger rate and
under reasonable assumptions on the binary lifetime and observability; 
see][]{volonteri09}.

The spectroscopic approach has an obvious drawback: a peculiar spectrum
does not guarantee the presence of a BHB in the nucleus of the
host. As an example, an unobscured BHB with both BHs active could
resamble the spectrum of a double peaked emitter \citep[see,
  e.g.,][]{eracleous1994}, where broad double--peaked lines are
emitted because of the almost edge--on, disk--like structure of the BL
region of a single BH.  A binary with a single accreting BH would show
a single shifted BL. If the shift corresponds to a relatively small
velocity along the line of sight ($\lsim 4\,000$ km s$^{-1}$), the same
signature could be emitted by a remnant of a binary coalescence,
recoiling because of anisotropic gravitational wave emission
\citep[e.g.][]{komossa08}\footnote{If the galaxy merger is gas rich,
  the maximum recoil velocity is expected to be $\lsim 100$ km s$^{-1}$
  \citep{bogdanovic07,dotti10b,volonteri10,kesden10}. In this case, a
  significant shift between the different sets of lines would not be
  compatible with a recoiling BH.}. Finally, both the cases can be
reproduced by a chance superposition of two AGN (or an AGN-galaxy
superposition) within the angular resolution of the used spectrograph
\citep[e.g.][]{heckman09}. The simplest way to discriminate between
these scenarios and the BHB hypothesis would be to look for a periodic 
oscillation of the BL shifts
around the host galaxy redshift. However, the orbital period of the
binary could be too long to be easily observed
\citep{begelman1980}. For these reasons other possible explanations
have been proposed for all the BHB candidates discussed in literature.

To overcome the paucity of BHB candidates and the uncertainties
related to their interpretation, we depict two ways:\\ {\it
  Theoretically} a better description of the spectrum of BHB is
needed, in order to identify other characteristic signatures of binaries. Few
attempt as been made up to date \citep[see, e.g.][]{bogdanovic08,
  bogdanovic09, shen10b, montuori2010}.\\ {\it Observationally } all
the BHB candidates lurking in large spectroscopic catalogue must be
identified, through a meticulous study of all the possible spectra, in
order to allow for follow-up studies on a significant sample of
objects.

In this paper we explore the second path, describing the results we
obtained from a comprehensive search of BHBs in the SDSS DR7. The code
we use, described in Section~\ref{sec_method}, automatically detects
sources with a spectrum consistent with a BHB, a double peaked
emitter, a superposition, or a recoiling BH. We present all the
peculiar objects we find in Section~\ref{sec_results}, where, for each
object, we compare our results with previous findings available in the
literature.
Conclusions are drawn in Section \ref{sec_conclusions}. 
Throughout the paper we will assume a standard cosmology with $H_0=70$ km 
s$^{-1}$ Mpc$^{-1}$, $\Omega_{\rm m}=0.3$ and $\Omega_{\Lambda}=0.7$.

\section{The method}\label{sec_method}

\subsection{Present analysis}

In this study we perform an automatic and systematic search for BHB
candidates in SDSS catalogue, looking for composite spectra of two
sources with a velocity difference up to 30\,000 km s$^{-1}$. More
specifically, we look for the presence of two sets of emission lines
(one broad and one narrow) with a small separation between them,
caused by the Keplerian rotation of one component of the binary
system. For this purpose we use the
method described in \citet{tsalmantza}. As a first step we 
extract a small set of components that can sufficiently reconstruct
the SDSS QSO spectra by using HMF
(Heteroscedastic Matrix Factorization or HMF), a bilinear model optimized with probabilistically 
justified weighted least-squares objective function, analogous to principal components analysis but
 making use of the observational noise model. 
The method uses a subset of data at rest-frame
wavelengths, described in detail in Section~(\ref{sample}), as a
training set to define the set of basis functions that minimize the
scalar $\chi_{\epsilon}^2$:

\begin{eqnarray}\label{e1}
\chi_{\epsilon}^2 \equiv \sum_{i=1}^N \sum_{j=1}^M
\frac{\left[f_{ij}-\sum_{k=1}^K a_{ik}
                      \,g_{kj}\right]^2}{\sigma^2_{ij}} \nonumber\\ 
+ \epsilon\,\sum_{k=1}^K \sum_{\ell=2}^{M}
 \left[g_{k\ell}-g_{k(\ell-1)}\right]^2
\quad ,
\end{eqnarray}

where the first term corresponds to the total $\chi^2$ of the fit of
the training data points $f_{ij}$ with errors $\sigma_{ij}$, by a set
of K components $g_{kj}$ and coefficients $a_{ik}$, over all the N
spectra and the M pixels of the training set. The second term
corresponds to a smoothing prior or regularization that prefers small pixel-to-pixel
variations in each resulting component. The strength of the smoothing
is set by the scalar $\epsilon$. An iterative procedure is used to
minimize $\chi_{\epsilon}^2$. In each step, we fix $g_{kj}$ and estimate the
optimal $a_{ik}$, and then hold the $a_{ik}$ fixed and estimate the
optimal $g_{kj}$. The procedure is repeated until the solution
converges. We initialize the fitting with the output of PCA when
applied to an extended set of training spectra.

We then use the resulting set of components to fit each observed spectrum 
at the redshift provided by SDSS. After that, we
repeat the fitting using {\it two} sets of components at different redshifts.  
Here one set is assumed to be at the SDSS
redshift, while the second is free to vary over a broad range of $z$,
corresponding to velocity differences up to 30\,000 km s$^{-1}$. If a second
redshift system is present, we expect the fit to significantly improve when we
add the second set of components. This procedure has
already been proven to successfully identify four of the known BHB
candidates \citep[J0927+2943, J1536+0441, J1050+3456, and J1000+2233, see][]{tsalmantza} with an
estimate of the velocity shifts between the two sets of lines 
  consistent with what obtained in previous studies.

\subsection{The training sample}\label{sample}

To train the HMF we use the same training set of quasar spectra that was used
in \citet{tsalmantza}. The sample consists of spectra in the
redshift range 0.1-1.5. However, since we are mainly interested in
detecting shifts between the narrow and the broad emission lines in
the QSO spectra, we check if our method is more sensitive in detecting
interesting objects when one of the two sets of components used to fit
each spectrum is representative of spectra with only narrow emission
lines. To define components that include only narrow lines, we use
galaxy spectra for the training of the HMF. The galaxy sample
consists of spectra classified spectroscopically as galaxies in SDSS
with S/N$>$20 and EW of the \Oii{} and \Oiii{} lines larger than 20
(10\,856 spectra).

All spectra used for the training and the application of the method
are derived from the 7th Data Release (DR7) of SDSS. Pixels with any
of the flags: SP\_MASK\_FULLREJECT, SP\_MASK\_BRIGHTSKY,
SP\_MASK\_NODATA, SP\_MASK\_NOSKY or pixels that correspond to zero
noise were treated as masked. All the spectra and their noise were
moved to the rest-frame (assuming $z=z_{\rm SDSS}$), resulting in
spectra with different spectral coverage. However, both HMF and PCA
(the output of which was used as an initialization to the HMF) require
common wavelengths for all the training spectra. The common wavelength range was
defined by the pixels that included information for at least 10 spectra of each type (i.e. galaxies or QSOs).
The final wavelength coverage was 1522.299-8352.183~\AA\ for
the QSOs and 3044.388-9193.905~\AA\ for the galaxies, corresponding to
7394 and 4801 pixels respectively. As a last step before the
application of the method all spectra were interpolated to common
wavelengths selected for each type of source using cubic splines. The
selected wavelengths are uniformly distributed in log space as in the
case of the original SDSS wavelengths. For the resulting spectra we
interpolated linearly the values of the masked pixels.

PCA was performed separately for the QSO and the galaxy
sample, using a number of spectra equal to the number of pixels selected for
each source. The training spectra were first projected into a subspace
orthogonal to the mean spectrum of the data set and the
flux in each spectral bin was divided with the RMS of the noise in
that bin, for all the non-masked pixels in the training sample. The
PCA results were used as an initialization to the HMF. The method was
run for a subset of approximately 1\,000 spectra of each type for a
different number of components and for 16 iterations, until
convergency. This test was also performed for four different values
(1,3,10 and 30) of the smoothing scalar $\epsilon$. To perform a simple cross validation, the resulting
components were also used to fit another subset of 1\,000 test spectra for
each type. Based on these tests, we decided to
use 14 and 7 components for the QSOs and galaxies and $\epsilon$
values of 10 and 1 respectively. Those numbers were also defined by
using the spectra of the known BHB candidates and testing the ability
of the components to detect them.

\subsection{Selection criteria}\label{selcrit}

We apply our newly developed fitting scheme to all the 54\,586 QSO
spectra of SDSS in the redshift range 0.1-1.5. Additionally, since one of the known BHB candidates
\citep[4C+22.25,][]{decarli_4c2225} is classified spectroscopically as
a galaxy in the SDSS catalog, we also applied the method to objects
with redshift from 0.2-1.5, that are classified by SDSS as
galaxies and have fiber magnitudes that correspond to $u-g<0.8$,
$g-r<1.5$, $r-i<1.0$ and $g<21$. We note that the majority of these sources 
(3\,518 out of 3\,929) were targeted as quasar candidates by SDSS.
For each spectrum the
fitting was performed for all the combinations of the extracted QSO and galaxy
components: i) two QSO components for both the spectral
components (hereafter QSO-QSO), ii) a QSO component for the spectrum
at the SDSS redshift and a galaxy components at the second redshift
(QSO-Galaxy) and iii) the opposite of case (ii) (Galaxy-QSO).
Out of the 175,545 fitting results we assigned
priority based on the following criteria:

\begin{itemize}
\item[{\it i-}] The fit significantly improves by adding a set of
  components at a second redshift, i.e. it corresponds to large
  $\chi^2$ difference. The threshold was set based on the values
  extracted for the four known candidates and by visual inspection of
  the fitting results for various $\chi^2$ difference values.

 \item[{\it ii-}] The peaks of the $\chi^2$ difference do not
   correspond to fits with unphysical properties (e.g. negative
   emission lines). Negative residuals are common when fitting
   the spectra with one set of components. Their strength, and
   therefore their impact in the results, can vary from very weak
   features usually caused by details in the continuum fitting and
   the presence of noise, to very strong features caused
   by poor fitting of emission lines. To exclude the latter
   cases (but not the former) from our final sample we examine only peaks of the
   $\chi^2$ differences that correspond to positive differences
   between the 99.5\% and the 0.5\% quartiles of the distribution of
   fluxes per pixel in each fitted spectrum
   (for each set of components).

\item[{\it iii-}] The peaks of the $\chi^2$ difference correspond to a
  difference in redshift between the two sources larger than 0.01.

\end{itemize}

In the case that Galaxy-QSO fitting was performed to the spectra,
there was a lot of contamination to our results due to two additional
reasons:

\begin{itemize}
\item[{\it i-}] In most of the cases the fitting of the narrow emission lines
  at the SDSS redshift by the galaxy components was resulting to unusual residuals of the broad
  lines, (e.g. in the case that no significant shift was present between the NLs and BLs),
  which were then fitted very poorly by the second set of
  components. To exclude these objects from our final sample we also
  measured the $\chi^2$ value of the fit by the second set of
  components restricted to an area of 500 \AA\ around the \Ha{} line and 200
  \AA\ around the \Hb{} line. Peaks corresponding to a very large
  value of $\chi^2$ at the areas of the BLs were excluded.

\item[{\it ii-}] In all cases the fitting of a QSO spectrum by the
  galaxy components will improve significantly when the QSO components
  are added to the fit. That is due to the fit of the broad emission
  lines by the second set of components. To make sure that the
  observed improvements were not only caused by this, we re-perform
  the fitting of the spectrum by two sets of QSO components, at the
  redshifts suggested by the Galaxy-QSO fitting.
\end{itemize}

\begin{figure}[h]
\includegraphics[angle=-90,width=0.99\columnwidth]{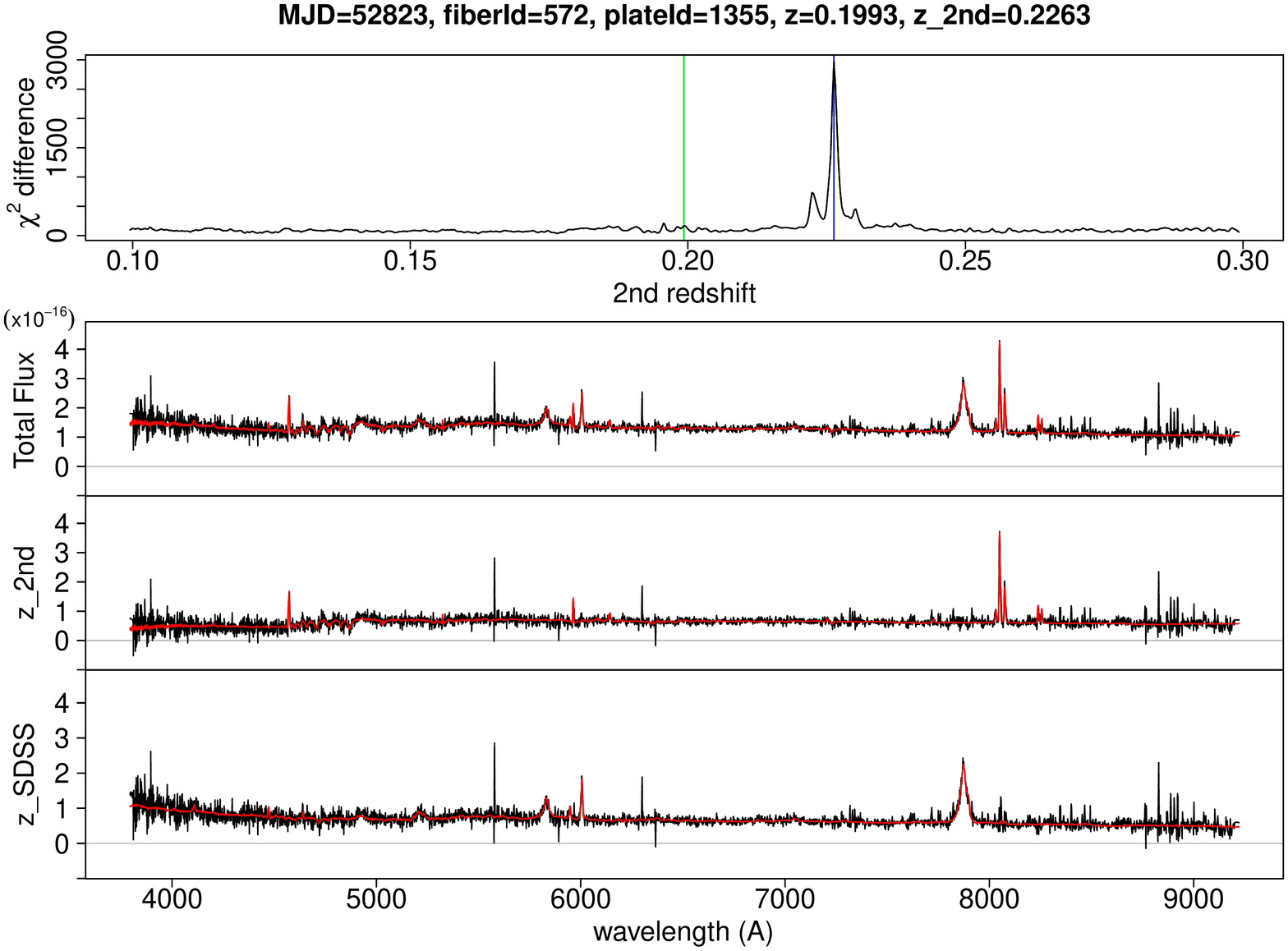}
\caption{An example of an interesting object (J1539+3333) detected by
  the method. \textbf{Top panel:} The $\chi^2$ difference between the
  fitting of the spectrum with 1 and 2 sets of components at different
  redshifts. The green and blue lines represent the SDSS redshift
  ($z_{\rm sdss}$) and the one with the largest $\chi^2$ difference
  ($z_{\rm fit}$). \textbf{2nd panel:} The fitting of the spectrum
  (black) with both sets of components (red).  \textbf{3rd panel:}
  Black line: residuals after subtracting the fit components $z_{\rm
    sdss}$. The fit with the set of components at $z_{\rm fit}$ is shown in red.
  \textbf{Bottom panel:} Black line:
  residuals after subtracting the fit at $z_{\rm fit}$. Red line: the fit with the set
  of components at $z_{\rm sdss}$. A color version of the plot is 
  available in the electronic version of the paper.}
\label{f1}
\end{figure}

The fits of the spectra selected using the above criteria were
visually inspected independently by two of the authors (PT and
RD). The fitting results showed that there is a lot of contamination,
e.g. from obvious Double Peaked Emitters (DPEs), misclassified objects in SDSS or spectra with wrong
estimated redshifts. By excluding all those cases, the most
interesting objects were identified. In Figure \ref{f1} we provide an
example of the fitting output for one of those sources.

The objects selected here are presented in detail in the following
section. We should point out that all the objects selected for their
interesting features are at redshifts below 0.8. For higher redshifts
the \Oiii{} line is not included in the spectra and the differences in
the fitting are detected mainly based on the \Oii{} line, which in
most of the cases is very faint or not well detected. For this reason
no reliable information for the narrow emission lines are included in
most of the spectra at higher redshifts.

\section{Results}\label{sec_results}

\begin{table*}
\caption{{\rm Summary of the peculiar objects found with our code. (1)
    Quasar name. (2-3) Right ascension and declination (J2000.).  (4)
    Modified Julian Date of the SDSS observation. (5) Plate. (6)
    Fiber. (7) Redshift of Narrow Lines. (8) Redshift of broad lines.
    (9) Method used to select the target. QG: Quasar--Galaxy; QQ -
    Quasar--Quasar; GQ: Galaxy--Quasar (see the text for the method
    description). (10) Classification: A- Asymmetric BL profile; B-
    Black hole binary candidate; D- Double-peaked emitter; E- Extreme
    double-peaked emitter; 
    O- Others (see section
    \ref{sec_results}).  }} \label{tab_candidates}
\begin{center}
\begin{tabular}{cccccccccc}
   \hline	      
    Obj.name    & R.A.    &  Dec.  & MJD & Plate & Fiber & $z_{\rm NL}$ & $z_{\rm BL}$ & Method & Class. \\
		&         &	   &     &       &       &              &              & 	&        \\
     (1)	&   (2)   &  (3)   & (4) &   (5) &  (6)  &  (7)         &  (8)         & (9)    & (10)   \\
   \hline 
J0012-1022 & 00:12:24.03 & -10:22:26.3 &52141&0651&072& 0.228 & 0.221 & QG,QQ	 & A   \\ % Ha   
J0155-0857 & 01:55:30.02 & -08:57:04.0 &52168&0665&597& 0.165 & 0.170 &    QQ	 & O   \\%S   \\ % Ha   
J0221+0101 & 02:21:13.15 & +01:01:02.9 &51869&0406&374& 0.354 & 0.364 & QG,QQ	 & O   \\%S   \\ % Ha   
J0829+2728 & 08:29:30.60 & +27:28:22.7 &52932&1267&066& 0.321 & 0.325 &       GQ & O   \\%S   \\ % Ha   
J0918+3156 & 09:18:33.82 & +31:56:21.2 &52990&1592&139& 0.452 & 0.457 &    QQ,GQ & D   \\ % M2   
J0919+1108 & 09:19:30.32 & +11:08:54.0 &53050&1740&399& 0.369 & 0.372 &    QQ	 & O   \\%S   \\ % Ha   
J0921+3835 & 09:21:16.13 & +38:35:37.6 &52731&1214&293& 0.187 & 0.182 &    QQ	 & A   \\ % Ha   
J0927+2943 & 09:27:12.65 & +29:43:44.1 &53389&1939&467& 0.713 & 0.698 & QG,QQ	 & B   \\ % Hb   
J0931+3204 & 09:31:39.05 & +32:04:00.2 &53386&1941&553& 0.226 & 0.226 &    QQ,GQ & O   \\ % Ha   
J0932+0318 & 09:32:01.60 & +03:18:58.7 &52254&0568&039& 0.420 & 0.401 &    QQ	 & B,D \\ % Hb   M2  0.414
J0936+5331 & 09:36:53.85 & +53:31:26.9 &52281&0768&473& 0.228 & 0.237 &    QQ	 & A   \\ % Ha   
J0942+0900 & 09:42:15.12 & +09:00:15.8 &52757&1305&281& 0.213 & 0.168 &       GQ & E   \\ % Ha   Har 0.290
J0946+0139 & 09:46:03.95 & +01:39:23.7 &51989&0480&480& 0.220 & 0.227 &    QQ	 & A   \\ % Ha   
J1000+2233 & 10:00:21.80 & +22:33:18.6 &53737&2298&102& 0.419 & 0.377 &       GQ & E,B \\ % Ha   
J1010+3725 & 10:10:34.28 & +37:25:14.8 &52993&1426&110& 0.282 & 0.276 & QG,QQ	 & O   \\ % O3   
J1012+2613 & 10:12:26.86 & +26:13:27.3 &53757&2347&513& 0.378 & 0.351 &    QQ	 & E,B \\ % Ha   
J1027+6050 & 10:27:38.54 & +60:50:16.5 &52375&0772&216& 0.332 & 0.301 &    QQ	 & E   \\ % Ha   
J1050+3456 & 10:50:41.36 & +34:56:31.4 &53431&2025&603& 0.272 & 0.258 &    QQ,GQ & B   \\ % Ha   
J1105+0414 & 11:05:39.64 & +04:14:48.2 &52356&0581&226& 0.436 & 0.406 &       GQ & E   \\ % Hb   M2 0.450
J1117+6741 & 11:17:13.91 & +67:41:22.7 &51942&0491&402& 0.248 & 0.253 &       GQ & O   \\%S   \\ % Ha   
J1154+0134 & 11:54:49.42 & +01:34:43.6 &52051&0515&099& 0.469 & 0.450 &    QQ	 & A,B \\ % Hb   
J1207+0604 & 12:07:55.83 & +06:04:02.8 &52376&0842&530& 0.136 & 0.128 &       GQ & O   \\%S   \\ % Ha   
J1211+4647 & 12:11:13.97 & +46:47:12.0 &53116&1449&001& 0.294 & 0.287 &    QQ,GQ & O   \\%S   \\ % Ha   
J1215+4146 & 12:15:22.78 & +41:46:21.0 &53120&1450&141& 0.196 & 0.206 &    QQ,GQ & O   \\ % Ha   
J1216+4159 & 12:16:09.60 & +41:59:28.4 &53120&1450&130& 0.242 & 0.233 &       GQ & O   \\%S   \\ % Ha   
J1328-0129 & 13:28:34.15 & -01:29:17.6 &52426&0911&333& 0.151 & 0.140 &    QQ,GQ & O   \\%S   \\ % Ha   
J1414+1658 & 14:14:42.03 & +16:58:07.2 &54523&2758&014& 0.237 & 0.242 &    QQ	 & O   \\%S   \\ % Ha   
J1440+3319 & 14:40:05.31 & +33:19:44.5 &53498&1646&283& 0.179 & 0.165 &       GQ & D   \\ % Ha   
J1536+0441 & 15:36:36.22 & +04:41:27.0 &54567&1836&270& 0.389 & 0.373 & QG,QQ	 & D,B \\ % Ha   
J1539+3333 & 15:39:08.09 & +33:33:28.0 &52823&1355&572& 0.226 & 0.199 & QG,QQ	 & O,B \\ % Ha   
J1652+3123 & 16:52:55.90 & +31:23:43.8 &52790&1343&593& 0.593 & 0.590 & QG,QQ,GQ & O   \\ % M2   
J1714+3327 & 17:14:48.51 & +33:27:38.3 &54591&2973&190& 0.181 & 0.186 &    QQ	 & O,B \\ % Ha   
\hline
 \end{tabular}
 \end{center}
\end{table*}

Our selection produced a list of 32 candidates of particular relevance
(see Table \ref{tab_candidates}). For each target, we re-analyzed
the SDSS spectrum, modeling it with a power law for the QSO
continuum emission, a host galaxy template at the redshift of the NLs
and a template of the iron complex, as described in \citet{decarli10a}
and \citet{derosa11}.  We fitted the broad components of \Mgii{},
\Hb{} and \Ha{} with 2 gaussian functions at the same peak. This
fitting approach aims to better constrain the peak wavelength, and is
not meant to reproduce the line profile in detail. Narrow lines are
masked when fitting the broad component; by construction of our
sample, there is a velocity offset between BLs and NLs. This
simplifies the measurement of the peak wavelengths of the two
components.

\begin{figure*}
\begin{center}
\includegraphics[width=\textwidth]{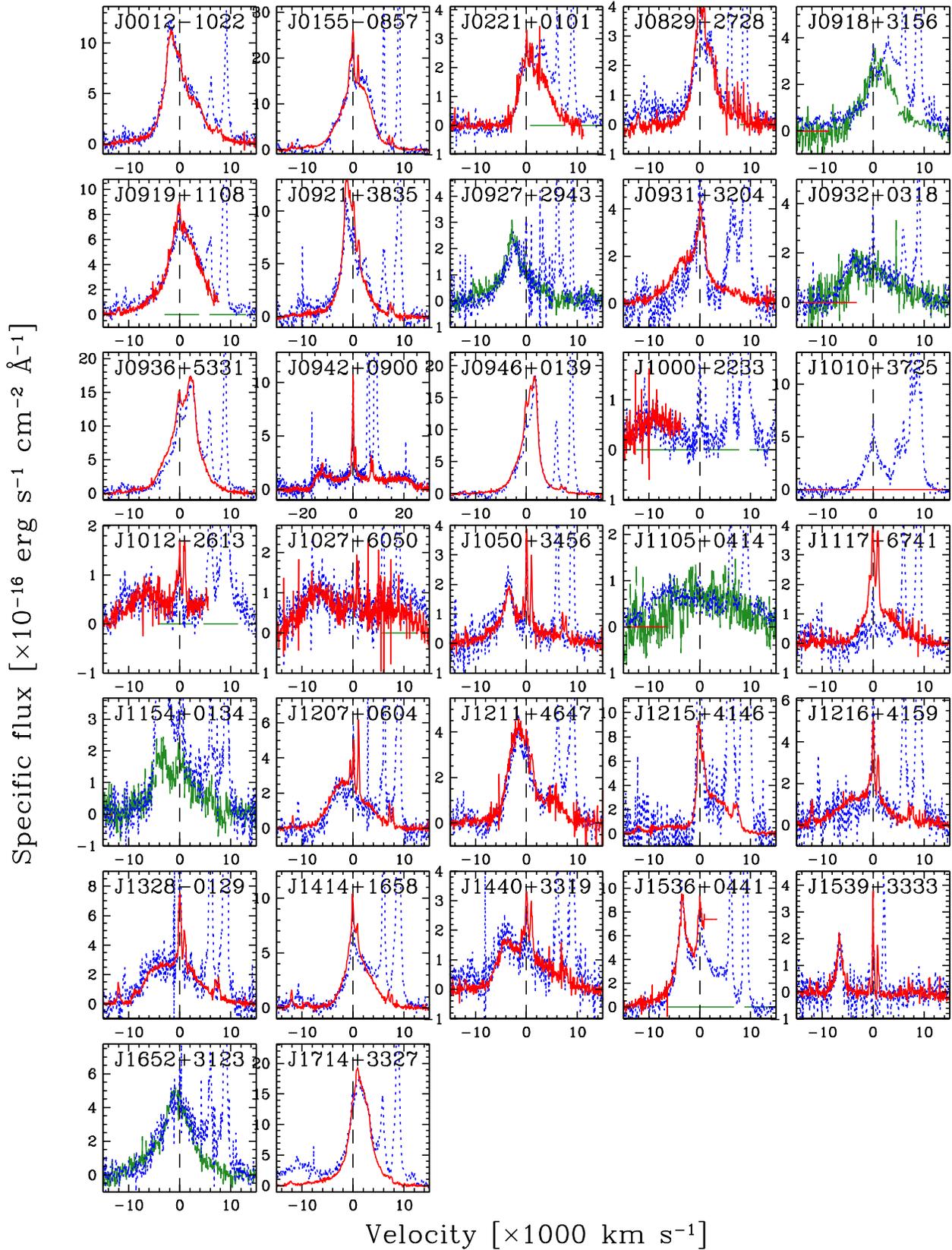}\\
\caption{Velocity diagrams of \Ha{} (red, solid lines), \Hb{} (blue,
  dotted lines) and \Mgii{} (green, dashed lines) for all our
  candidates. The flux of \Hb{} is scaled up to match the one of \Ha{}
  or \Mgii{}. }\label{fig_all}
\end{center}
\end{figure*}

Peak wavelengths are then converted into velocity shifts:
\begin{equation}
v_{\rm BL} = c \frac{z_{\rm BL} - z_{\rm NL}}{1 + z_{\rm NL}}.
\end{equation}
Continuum-subtracted velocity plots of all the interesting targets are
shown in Figure \ref{fig_all}.

A rough classification scheme was set according to: 1) the magnitude
of the velocity shift, in particular when comparing \Mgii{} and Balmer
lines; 2) the presence of strong asymmetries in the line profiles; 3)
the occurrence of secondary bumps or peaks; 4) additional information
from other emission lines or from the SDSS images. We define five
classes of objects, namely; 
$i$) {\bf black hole binary candidates}, which are expected to show similar 
velocity shifts for all the BLs, and a variety of line profiles 
\citep[e.g.,][]{shen10b}.
$ii$) quasars with {\bf asymmetric line profiles}, with small 
($\lsim2\,000$ \kms{}) shifts of BL peaks. These features
are observed in some ``normal'' type-I
AGN, and they are possibly related to asymmetries in the BL region
\citep[e.g.,][]{bentz10}.
They may also be associated to a velocity-dependent 
Balmer decrement of broad lines. 
$iii$) {\bf double peaked emitters} (DPEs), characterized
by symmetric features in line profiles (e.g., a secondary peak in the 
red wing of the line, at the opposite velocity with respect to a 
blue-shifted peak). Another property of DPEs is that different lines 
(in particular low- and high-ionization lines) may show very different 
shapes and shifts \citep{halpern96}. These properties are usually associated
to a disk-like structure of the BL region \citep{eracleous1994}. 
$iv$) {\bf extreme double peaked emitters} (see below).
$v$) {\bf others}, i.e. objects with small shifts or poor signal-to-noise
spectra, preventing us from a clear interpretation, or objects with
very peculiar properties, not belonging to any of the aforementioned classes.

Note that all the BHB candidates with small velocity shifts could also
be recoiling BHs, though they are expected to be rarer than binaries
\citep{dotti09,volonteri10}\footnote{Note that these estimates could
  be affected by our incomplete understanding of the orbital decay of
  binaries at subparsec scales \citep[see, e.g.][]{colpi2009,lodato09}.}.
In the following, we will not distinguish between these two
cases, including both in the ``BHB candidates'' class.

This classification produced 9 BHB candidates, including the 5 known
candidates. For the new 4 sources, other interpretations are also
plausible, including an extremely rare case of quasar--galaxy superposition 
for one of them. Five quasars show very high velocity shifts ($\gsim 5\,000$
\kms) and relatively faint lines. These objects probably represent
extreme cases of DPEs (hereafter, they will be referred to as
EDPEs). In the following, we discuss the properties of each source
individually, reporting our interpretation on the nature of the
object.

\begin{center}
{\it J0012-1022}
\end{center}

The Balmer broad lines of this $z_{\rm NL}=0.228$ quasar show a peak
$\sim 1\,700$ \kms{} blue-shifted with respect to narrow lines.  The
line profile is clearly asymmetric. \Ha{} and \Hb{} have identical
profiles, with $F($\Ha$)=2.6 F($\Hb$)$. A bump in the red wing of
Balmer lines suggests that this is a strongly asymmetric double-peaked
emitter \citep[see also][]{strateva03,shen10a}, though \citet{shen10b}
showed that the line profile of this source can be ascribed to an
unequal mass BHB. {\bf Classified: asymmetric line profile.}

\begin{center}
{\it J0155-0857}
\end{center}

The \Ha{} line of this source shows a small ($\sim1\,500$ \kms{})
red-shift with respect to the narrow lines. A slight asymmetry in the
line profile is reported. The asymmetry is clearer in the \Hb{}
profile, which peaks at longer wavelengths ($\Delta v \approx2\,200$
\kms). \citet{shen10a} reported a blue-shift of $\sim600$ \kms{} for
\Ha{} and $\sim2\,300$ \kms{} for \Hb{}.  The relatively small velocity
shift and the difference in the profiles of Balmer lines suggest that
this is a normal quasar. {\bf Classified: others.}

\begin{center}
{\it J0221+0101}
\end{center}

Both \Ha{} and \Hb{} broad components of this $z_{\rm NL}=0.354$
quasar show a $\sim 1\,300$ \kms{} shift with respect to the narrow
lines.  Line profiles are rather boxy with no obvious asymmetry.
\citet{shen10a} report no significant shift for \Ha{}, and an
exceedingly pronounced shift for \Hb{} ($\sim3\,100$ \kms{}). 
{\bf Classified: others.}

\begin{center}
{\it J0827+2728}
\end{center}

The Balmer lines of this quasar show a small ($\sim900$ \kms{}) 
red-shift with respect to the NLs, which may also be consistent
with a strongly asymmetric line profile. {\bf Classified: others.}

\begin{center}
{\it J0918+3156}
\end{center}

The peculiar properties of this object were first reported by
\citet{bonning07}. The \Mgii{} and \Hb{} broad lines are red-shifted
with respect to the NLs. However, the shift is $\sim3\,000$ \kms{} for
\Hb{} and only $\sim1050$ \kms{} for \Mgii{}
\citep[see][]{bonning07,shen10a}, suggesting that the shift is due
neither to a BHB nor a recoil. {\bf Classified: double peaked
  emitter.}

\begin{center}
{\it J0919+1108}
\end{center}

The broad emission lines of this source are slightly red-shifted
($700-1\,000$ \kms{}, depending on the subtraction of the \Nii{}
lines). The SDSS image of the quasar reveals a complex morphology,
probably resulting by a strong gravitational interaction or a merger
with a nearby galaxy. \citet{shen10a} reported a blue-shift for the
broad component of \Ha{} and a red-shift for \Hb{}, which is not 
confirmed in our analysis. {\bf Classified: others.}

\begin{center}
{\it J0921+3835}
\end{center}

This object shows clearly asymmetric Balmer lines, with a peak
at $\sim1\,200$ \kms{} blue-wards \citep[consistent with the values
reported by][]{shen10a}. The \Ha{}/\Hb{} flux ratio
is $\sim3.3$. An unidentified line is observed at 5812 \AA{}, 
corresponding to a $\sim 2\,300$ \kms{} shift with respect to 
the \Hb{} rest frame. A similar peak is not observed in the 
\Ha{} profile. {\bf Classified: asymmetric line profile.}

\begin{center}
{\it J0927+2943}
\end{center}

One of the known BHB candidates \citep{bogdanovic09,dotti09}. This
quasar was first reported by \citet{komossa08} as a recoiling BH
candidate. Three sets of lines are observed, two narrow systems at
$z=0.713$ and $z=0.699$ and a broad line system consistent with the
second set of narrow lines. The shift between the two redshifts is
$\sim2\,600$ \kms{}. This object has been interpreted as a chance
superposition in a galaxy cluster by \citet{heckman09} and
\citet{shields2009}. However, later observational follow-ups have
disproved the presence of a galaxy cluster in the source field
\citep{decarli09s}. {\bf Classified: black hole binary candidate.}

\begin{center}
{\it J0931+3204}
\end{center}

The spectrum of this quasar shows relatively broad and asymmetric
narrow lines, which makes the determination of $z_{\rm NL}$
uncertain. The blue wing of \Ha{} shows a bump at $\sim 3\,500$ \kms{}
with respect to the peak of the line. This feature is not observed in
\Hb{}. This object was not reported in previous studies of quasars
with peculiar NL profiles \citep[e.g.][]{wang09,liu10a,liu10b}.
\citet{shen10a} flagged this source as a DPE candidate. {\bf
  Classified: others.}

\begin{center}
{\it J0932+0318}
\end{center}

This source shows a shift of $\sim4\,000$ \kms{} in the peak of the
\Mgii{} and the \Hb{} lines. \citet{shen10a} reported two different
velocity offsets for \Hb{} ($\sim3\,900$ \kms) and \Mgii{} ($\sim1\,100$
\kms) respectively, but such a difference is not confirmed by our
analysis. The \Hb{} line appears slightly asymmetric, though the line
profile is poorly determined.  This object has been studied in detail
by \citet{barrows11}, who suggested it to be a DPE from an asymmetric
BL region, or a BHB. {\bf Classified: double peaked emitter} or
{\bf  black hole binary candidate.}

\begin{center}
{\it J0936+5331}
\end{center}

The Balmer lines of this $z_{\rm NL}=0.228$ quasar show a strongly
asymmetric profile, with a bright blue peak at $2\,100$ \kms{}
\citep[$2250$ \kms{} in the analysis by][]{shen10a} with respect to
the narrow lines. The red wing of \Ha{} is $3.1\times$ brighter than
\Hb; the blue wing is $4.0\times$ brighter. No peak or bump is
observed in the blue wing at $2\,100$ \kms{}. This source was listed in
the ``DPE Auxiliary sample'' by \citet{strateva03}. {\bf Classified:
  asymmetric line profile.}

\begin{center}
{\it J0942+0900}
\end{center}

This object has been labeled as a galaxy by the SDSS automatic
pipeline, presumably because the broad lines are too flat and
extended. The Balmer lines are $\sim38\,000$ \kms{} broad (the broadest
ever reported!) and asymmetric
(the blue peak is brighter and peaks around $\sim -11\,500$ \kms{},
while the red side is fainter, flatter and extends red-wards of
$+20\,000$ \kms{}).  The flux ratio between \Ha{} and \Hb{} is 2.6, and
the lines have similar profile.  This object is the most extreme DPE
ever discovered in terms of line width.  This object may also
represent the prototypal to explain the peculiar features of other
(less extreme and fainter) cases (e.g., J1000+2233, J1012+2613, etc).
{\bf Classified: extreme double peaked emitter.}

\begin{center}
{\it J0946+0139}
\end{center}

The Balmer lines of this $z_{\rm NL}=0.2203$ quasar peak at 1550
\kms{} blue-wards of the NL system \citep[see
  also][]{shen10a}. \citet{boroson10} reported an ``anomalous \Hb{}
profile'' for this source. The broad lines show no obvious second
peak, but the blue wings are slightly more extended than the red
ones. Also, the flux ratio between \Ha{} and \Hb{} is larger in the
blue wing. {\bf Classified: asymmetric line profile.}

\begin{center}
{\it J1000+2233}
\end{center}

This source was serendipitously discovered by our group
\citep{decarli_4c2225} out of the SDSS database, and share some of the
properties of J0942+0900.  The Balmer lines appear faint and extremely
blue-shifted ($\sim 8\,700$ \kms{}).  Also the \Mgii{} line shows a
blue-shift, but its magnitude is poorly constrained since the line is
only partially covered by the SDSS spectrum.  This quasar was labeled
as a galaxy by the SDSS pipeline. \citet{shen10a} reported
inconsistent velocity offsets for \Hb{} ($4\,500$ \kms{} {\it
  red}-wards) and \Mgii{} ($2\,700$ \kms{} blue-wards).  {\bf
  Classified: extreme double peaked emitter} or {\bf black hole binary
  candidate.}

\begin{center}
{\it J1010+3725}
\end{center}

This object shows a complex \Oiii{} profile (both for the 4959 and the
5\,008 \AA{} emission lines). Two peaks are observed, with a velocity
difference of $\sim 1\,400$ \kms{}.  The blue peaks are fainter. Other
narrow emission lines (\Hb{}, \Ha{}, \Sii{}) appear normal. This
source was labeled as ``anomalous \Oiii{} profile'' by
\citet{boroson10}.  A peculiar \Oiii{} profile was also reported by
\citet{shen10a}. This quasar was not included in previous studies
on double-peaked NL objects
\citep[e.g.][]{wang09,liu10a,liu10b}. {\bf Classified: others.}

\begin{center}
{\it J1012+2613}
\end{center}

The Balmer lines of this $z_{\rm NL}=0.3783$ quasar show similar
properties to those of J1000+2233: The BLs peak $\sim 6\,000$ \kms{}
blue-wards of the NLs. The red wing of \Ha{} and the blue wing of
\Mgii{} are not covered. {\bf Classified: extreme double peaked
  emitter} or {\bf black hole binary candidate.}

\begin{center}
{\it J1027+6050}
\end{center}

This is the only EDPE already known \citep{strateva03}. The blue peaks
of \Ha{} and \Hb{} are 7\,000 \kms{} blue-shifted with respect to
NLs. The shift is missed in the analysis by \citet{shen10a}, probably
because of the faintness of the lines. {\bf Classified: extreme double
  peaked emitter.}

\begin{center}
{\it J1050+3456}
\end{center}

This source was discovered by \citet{shields09} out of the SDSS
database. The broad component of Balmer lines is clearly shifted
\citep[$\sim 3\,400$ \kms{} blue-wards; a similar value was found
  by][]{shen10a}.  No NL is observed at the redshift of the BLs. {\bf
  Classified: black hole binary candidate.}

\begin{center}
{\it J1105+0414}
\end{center}

This quasar was found in the galaxy sample. The broad
\Hb{} line peaks $\sim 6\,000$ \kms{} blue-wards of the narrow
component. Due to its faintness, the line profile is poorly
constrained. The \Mgii{} line is equally faint, but no 
obvious shift is observed, supporting the DPE interpretation
for this source. {\bf Classified: extreme double
  peaked emitter.}

\begin{center}
{\it J1117+6741}
\end{center}

The \Ha{} broad component of this source appears slightly redshifted
with respect to the NLs. The properties of \Hb{} are difficult to
characterize, due to its intrinsic faintness. {\bf Classified:
  others.}

\begin{center}
{\it J1154+0134}
\end{center}

The \Hb{} and \Mgii{} lines of this $z_{\rm NL}=0.469$ quasar have
identical profiles, with a peak $\sim 3\,500$ \kms{} blue-wards
of the expected wavelengths and a rather broad red wing.
The line profiles resemble the one of other sources with
asymmetric lines (e.g., J0012-1022), but the magnitude
of the shift and the similarity between \Hb{} and \Mgii{}
support the BHB hypothesis. The low S/N of the spectrum 
of the SDSS spectrum hinder any conclusion on the nature
of this source. {\bf Classified: asymmetric line profile} or
{\bf black hole binary candidate.}

\begin{center}
{\it J1207+0604}
\end{center}

The broad component of the Balmer lines of this quasar is rather
symmetric but shifted $\sim 2\,500$ \kms{} blue-wards with respect to
NLs. The flux ratio between \Ha{} and \Hb{} is $\sim 4$, roughly
constant with respect to the line-of-sight velocity. This source was
not included in the compilation by \citet{shen10a}.  {\bf Classified:
  others.}

\begin{center}
{\it J1211+4647}
\end{center}

The bulk of the Balmer line broad components of this source is 
shifted $\sim1\,700$ \kms{} blue-wards of the NL system
\citep[consistent values are reported in][]{shen10a}. The 
\Ha/\Hb{} flux ratio is $\sim4$, constant along the velocity
profile. The \Ha{} line shows a bump at $\sim 6\,000$ \kms{}
in the red wing, possibly revealing the DPE-like nature
of this source. Such a feature is not clearly observed for 
\Hb{} because of the superposition of the \Oiii{} doublet. {\bf
  Classified: others.}

\begin{center}
{\it J1215+4146}
\end{center}

This quasar shows a peculiar Balmer line profile.  The bulk of \Ha{}
emission arises from a bright bump in the red wing. The blue side of
the line may also present a faint wing, the actual presence of which
depends on the continuum modeling. At zero order, the \Hb{} line shows
analogous profile. However, the feature in the red wing is $\sim 11$
times fainter than what observed in \Ha{}.  The interpretation of this
object is unclear.  \citet{boroson10} labeled this source as a `no
broad line' quasar. {\bf Classified: others.}

\begin{center}
{\it J1216+4159}
\end{center}

The \Ha{} broad emission of this quasar is clearly blue-shifted ($\sim
2\,300$ \kms{}). The line profile shows no relevant asymmetry. The \Hb{}
broad component is barely detected, its flux being $\sim7$ times
fainter than \Ha{}. {\bf Classified: others.}

\begin{center}
{\it J1328-0129}
\end{center}

The broad component of the Balmer lines in this $z_{\rm NL}=0.1514$
quasar are blue-shifted ($\sim 3\,100$ \kms{}) with respect to NLs.
The line profile is boxy, with no significant asymmetry. The
\Ha/\Hb{} flux ratio is $\sim 5$, constant over the velocity 
range. This object was not included in the analysis by \citet{shen10a}.
\citet{strateva03} and \citet{bian07} listed this source as
a DPE. {\bf Classified: others.}

\begin{center}
{\it J1414+1658}
\end{center}

The bulk of the BLs of this quasars is redshifted ($\sim 1\,200$ \kms)
with respect to NLs. The red wing is brighter. The \Ha{}/\Hb{} flux
ratio is $\sim4$ in the blue wing and around 3 in the red wing.  This
object was labeled as a DPE candidate by \citet{shen10a}.  
{\bf Classified: others.}

\begin{center}
{\it J1440+3319}
\end{center}

The \Ha{} line of this source peaks at $\sim 3\,700$ \kms{} blue-wards
of the NLs, and shows an extended red wing. The \Hb{} line profile
is similar. The properties of this quasar are half the way
between the objects with asymmetric line profiles (e.g., J1154+0134)
and the typical DPEs, though this source has not been included in any 
previous compilation of DPEs.  {\bf Classified: double peaked
  emitter. }

\begin{center}
{\it J1536+0441}
\end{center}
The peculiar properties of this object were first reported by 
\citet{boroson09}. The broad lines show two peaks, one consistent
with the rest-frame of the galaxy as set by NLs, the other 
significantly blue-shifted ($\sim3\,400$ \kms). \citet{boroson09}
proposed the BHB interpretation for this source. However, following
observations covering the red wing of \Ha{} revealed the presence
of a bump in the line wing \citep{chornock10}, a feature commonly observed in DPEs. {\bf
  Classified: double peaked emitter} or {\bf black hole binary
  candidate.}

\begin{center}
{\it J1539+3333}
\end{center}

From the spectroscopic point of view, the properties of this source
are similar to those of J0927+2943.  The spectrum presents three sets
of lines at two different redshifts: Broad balmer lines (driving the
redshift estimate by the SDSS pipeline) and faint narrow lines are
detected at $z_1=0.1993$. Another set of (brighter) narrow lines is
observed at $z_2=0.2263$. The corresponding velocity shift is $\sim
6\,600$ \kms{}. A careful inspection of the SDSS image of this
source reveals an extended stellar wing South-wards of the quasar.
If this belongs to the quasar host galaxy, then it would reveal a
strongly perturbed morphology. On the other hand, it could be
that this is a superposed galaxy. In this case, since $z_{\rm BL}<z_2$,
the galaxy would be in the background of the quasar.
This scenario is usually extremely unlikely, given the high velocity 
differences \citep{dotti10}. However, the SDSS image reveals the presence
of a rich galaxy cluster South-West of the quasar, which may enhance the 
galaxy density on the sky plain by few orders of magnitudes.
Follow-up observations aimed at directly measuring the redshift of 
the stellar wing are needed to fully understand the nature of this source.
{\bf Classified: black hole binary candidate} or {\bf others}.

\begin{center}
{\it J1652+3123}
\end{center}

The \Hb{} and \Mgii{} broad lines of this $z_{\rm NL}=0.5929$ quasar
show a small blue-shift ($\sim 500$ \kms) with respect to NLs. The
line profiles are similar and do not show any significant asymmetry.
Our shift estimates are consistent with those reported by \citet{shen10a}.
Given the small velocity difference, this source is probably a normal
quasar \citep{bonning07}. {\bf Classified: others.}

\begin{center}
{\it J1714+3327}
\end{center}

The Balmer lines of this source show a clear red-shift 
\citep[$\sim 1\,300$ \kms, consistent with the values reported in][]{shen10a}.  
The line profiles are symmetric. The \Ha/\Hb{} flux ratio is around
3. {\bf Classified: others} or {\bf black hole binary candidate.}

\section{Summary \& discussion}\label{sec_conclusions}

We presented the outcome of our automatic and systematic search for
massive BHBs. We have found 9 BHB candidates in the
SDSS DR7. Of these, 5 have already been extensively discussed in
literature.  The 4 new candidates are J1012+2613, J1154+0134, J1539+3333
and J1714+3327. For each one of them a BHB is not the
only possible explanation: The peculiar spectrum of J1012+2613 can be
explained also as an extreme case of double peaked emitter; J1154+0134
has a too noisy spectrum to exclude other explanations; J1539+3333
may be a rare superposition of a quasar and a galaxy; The small
shift between broad and narrow lines in J1714+332 ($\approx 1\,300$ km
s$^{-1}$) does not necessarily imply the presence of a BHB. A more
detailed understanding of the expected spectral features of BHBs and
observational follow-ups are needed to confirm or dismiss the BHB
hypothesis for all the 9 candidates presented here.

Our method also automatically detected a number of
other interesting objects with peculiar spectral features:

{\it i-} 4 objects show strong asymmetries in the line
  profiles, with a peak offset $\gsim 2\,000$ \kms{} (either red- or
  blue-wards) and a longer wing in the opposite velocity range with no
  secondary peak.

{\it ii-} 3 objects have BL properties analogous to what
  typically observed in DPEs, even if the secondary peak is not
  prominent. None of them appeared in the compilation by
  \citet{strateva03}.

{\it iii-} We provide strong evidence of a new class of extreme
  double-peaked emitters, with very broad (FWHM$>$10\,000 \kms) and
  rather faint emission lines. The main peak of these lines show huge
  velocity shifts ($>5\,000$ \kms) with respect to the NLs. For a
  comparison, only 5 objects out of 138 in \citet{strateva03} have a
  shift of the brighter peak of \Ha{} larger than 5\,000 \kms, and none
  of them exceed 7\,000 \kms{}. Note that the ``extreme double-peaked
  emitter'' explanation is possible also for one of the BHB candidates
  already discussed in literature \citep[J1000+2233,][]{decarli_4c2225}.

{\it iv-} out of the remaining 13 quasars, 12 show clear yet
  relatively small (500--3\,000 \kms) velocity offsets between NLs and
  BLs, and no obvious asymmetries in the line profiles of the broad
  components.

{\it v-} the case of J1539+3333 is worth of specific discussion. This quasar 
has few spectral features in common
with the BHB candidate J0927+2943, with a set of broad and narrow
emission lines shifted with respect to a brighter set of narrow
emission lines. However, the velocity shift is extremely high
($\approx 6\,600$ \kms). As a consequence, this object cannot be
explained in terms of a recoiling BH, since the maximum kick velocity
has been constrained with fully GR simulations to be $\lsim 4\,000$ \kms
\citep{baker07, herrmann07, campanelli07, schnittman07, lousto09,
  vanmeter10}. A natural explanation can be a superposition of an AGN
and a galaxy within a galaxy cluster. Such model has been ruled out
for J0927+2943 because no galaxy cluster was observable in the field
\citep{decarli09s}. For J1539+3333 a superposition is not ruled out by
observations. The redshift of this source is $z\approx 0.2$, close to
the theoretically estimated peak of superpositions in clusters
\citep{dotti10}. However, the line shift corresponds to a relative
velocity between the two galaxies $\gsim 1.5$ times larger than the
maximum relative velocity theoretically expected \citep{dotti10}. 
A better comprehension of this peculiar object
needs a more detailed study of its field, as already performed for
J0927+2943 \citep{decarli09s} and J1536+0441
\citep{decarli09,wrobel09}.

\section*{Acknowledgments}

The authors would like to thank C. A. L. Bailer-Jones and F. Walter for useful discussions and support. DWH was partially supported by the NSF (grant AST-0908357) and a research fellowship from the Alexander von Humboldt Foundation.

Funding for the Sloan Digital Sky Survey (SDSS) has been provided by the Alfred P. Sloan Foundation, the Participating Institutions, the
National Aeronautics and Space Administration, the National Science Foundation, the U.S. Department of Energy, the Japanese Monbukagakusho, and the Max Planck Society. The SDSS Web site is http://www.sdss.org/. The SDSS is managed by the Astrophysical
Research Consortium (ARC) for the Participating Institutions. The Participating Institutions are The University of Chicago, Fermilab,
the Institute for Advanced Study, the Japan Participation Group, The Johns Hopkins University, the Korean Scientist Group, Los Alamos
National Laboratory, the Max-Planck-Institute for Astronomy (MPIA), the Max-Planck-Institute for Astrophysics (MPA), New Mexico State
University, University of Pittsburgh, University of Portsmouth, Princeton University, the United States Naval Observatory, and the
University of Washington.

\label{lastpage}

\end{document}